# Polyp Segmentation in Colonoscopy Images Using Fully Convolutional Network

Mojtaba Akbari, Majid Mohrekesh, Ebrahim Nasr-Esfahani, S.M. Reza Soroushmehr, Nader Karimi, Shadrokh Samavi, Kayvan Najarian

*Abstract—* Colorectal cancer is a one of the highest causes of cancer-related death, especially in men. Polyps are one of the main causes of colorectal cancer and early diagnosis of polyps by colonoscopy could result in successful treatment. Diagnosis of polyps in colonoscopy videos is a challenging task due to variations in the size and shape of polyps. In this paper we proposed a polyp segmentation method based on convolutional neural network. Performance of the method is enhanced by two strategies. First, we perform a novel image patch selection method in the training phase of the network. Second, in the test phase, we perform an effective post processing on the probability map that is produced by the network. Evaluation of the proposed method using the CVC-ColonDB database shows that our proposed method achieves more accurate results in comparison with previous colonoscopy video-segmentation methods.

## I. INTRODUCTION

Colorectal cancer was the second cause of death in the U.S. in 2015 [1] and the third in 2017 [2]. Segmentation and detection of cancer in early stages of the disease will increase the chance of survival. Colonoscopy is the preferred method for analyzing inside the colon and also removing colorectal polyps. Polyps are the most important cause of the colorectal cancer. Colonoscopy is an operator-based method and human mistakes and also lack of sensitivity increase the need for computer-aided methods to segment these polyps in colonoscopy videos. Segmentation of colorectal polyp is also a challenging task because of variations in shape and color intensity of polyps in colonoscopy frames. Different methods have been proposed with the aim of accurate segmentation. We categorize research work in polyp segmentation into three main approaches. The first approach belongs to those image processing based segmentations which do not use any learning methods. The second group of approaches belongs to those methods which first extract features and then use classifiers for segmentation. In the third category we group those approaches that use convolutional neuronal networks (CNN) and perform the segmentation.

In this paper we propose a novel polyp segmentation method based on cascading of CNN. A smart patch selection method enhances the performance of the CNN. An adaptive thresholding is used, and then largest connected component is selected to further enhance the accuracy of our segmentation method.

In Section II of this paper some related existing research work are reviewed. In Section III, we present our proposed CNN structure and patch selection method for training FCN-8S. In Section IV we evaluate our proposed method with CVC-ColonDB database [18]. Concluding remarks are presented in Section V of the paper.

## II. RELATED WORKS

The first approach in polyp detection is to use image processing segmentation methods. Many methods have been proposed to segment the polyps automatically. A method proposed in [3] based on some information called "*image depth of valleys*" to segment colorectal polyps. In this method the watershed algorithm is used to segment images into polyp candidate regions and then classifies each region into polyp and non-polyp. This classification is based on regions information and "*depth of valleys*" in each region. Region information contains mean and standard deviation of each region and depth of valleys is based on calculation of eigenvalues and eigenvectors of the gradient image. Ganz *et al.* [4] propose a method based on Hough transform to detect region of interest (ROI) and specular reflection suppression with exemplar-based image inpainting as a preprocessing method. Then, they use a method using ultrametric contour map (UCM), called shape-UCM [5] for image segmentation. Shape-UCM works based on image gradient contours and spectral clustering. After performing shape-UCM algorithm, they use a scheme to improve edges resulted from the shape-UCM algorithm. The method of [4] works on the LAB color space and uses the image texture as a feature to refine edges. To overcome false positive in the resulting map, ellipse fitting algorithm is used to extract polyp boundaries from all candidate boundaries and regions. Method of [6] uses an improved watershed algorithm, named "*marker-controlled watershed*" method, as the initial stage for segmenting polyps. Authors of [6] also use the region-maxima method for selection of an initial point in the watershed algorithm and then they use elliptical fitting to discard unwanted regions resulted in the previous step.

The second approach in polyp detection is feature extraction from image patches and labeling of patches as

Mojtaba Akbari, Majid Mohrekesh, Ebrahim NasrEsfahani, and Nader Karimi, are with the Department of Electrical and Computer Engineering, Isfahan University of Technology, Isfahan 84156-83111, Iran.

S.M. Reza Soroushmehr is with the Department of Computational Medicine and Bioinformatics and Michigan Center for Integrative Research in Critical Care, University of Michigan, Ann Arbor, U.S.A.

Shadrokh Samavi is with the Department of Electrical and Computer Engineering, Isfahan University of Technology, Isfahan 84156-83111, Iran. He is also with the Department of Emergency Medicine, University of Michigan, Ann Arbor, U.S.A.

Kayvan Najarian is with the Department of Computational Medicine and Bioinformatics, Department of Emergency Medicine and the Michigan Center for Integrative Research in Critical Care, University of Michigan, Ann Arbor, U.S.A.

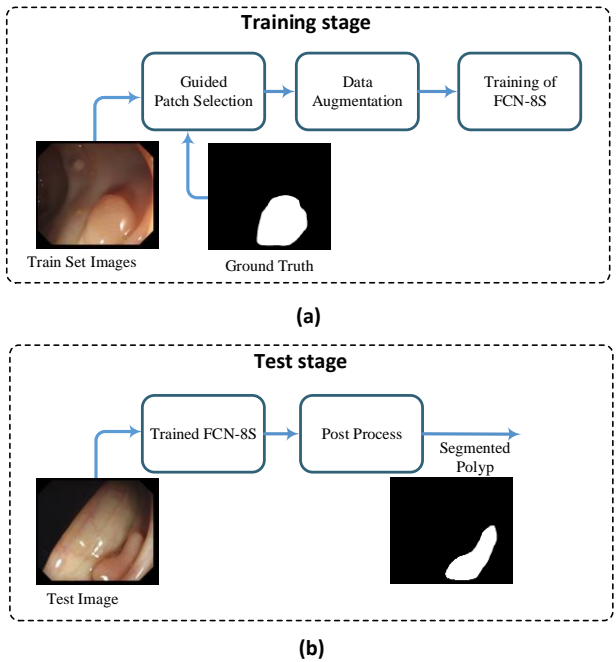

Figure 1. Block diagram of the proposed segmentation of colonoscopy polyps, (a) training phase, (b) test phase.

polyp and non-polyp based on extracted features. Tajbakhsh *et al.* [7] propose a method based on the work of [8] and Canny edge detector in each of the three RGB channels. This is done to produce edge maps and then the algorithm extracts oriented patches for each pixel to classify them as polyp or non-polyp. Proposed feature extraction method of [7] extracts sub-patch with 50% overlap and calculates their average vertically resulting one dimensional signal and then uses DCT coefficients as a feature for each extracted patch. Finally, it uses two-stage random forest classifier to label each patch. The first stage classifier converts low-level features into high level features and feed them to the second stage of the classifier to classify each patch into polyp and non-polyp classes.

The third approach for polyp detection is using Convolutional Neural Network (CNN) for segmentation of polyps. CNN is a type of deep learning method where trainable filters and pooling operations are applied on raw images to extract complex features [9]. CNN has been used to extract features to be fed into a subsequent classifier. Tajbakhsh *et al.* [10] analyze CNN results to see whether a full training, or a fine tuning method, works better in medical applications such as colonoscopy polyp segmentation. They showed that fine tuning works better than full training. In [11] CNN is used as a feature extractor in three scales patch representation for polyp segmentation. CNN calculates 60 features for each input patch, then uses fully-connected layer with 256 neurons for classification of each input patch. Moreover, Gaussian filter is employed to smooth the segmentation results and decrease noise after performing CNN. The method proposed in [12] uses three convolution layers and two pooling layers for extracting features from RGB patches and fully-connected layer for classification of 1024 extracted features.

New generation of CNNs uses deconvolution layers for generating probability map in image segmentation tasks. This newer approach is achieved by replacing fully-connected layer with deconvolution and using the information of previous layers for increasing segmentation accuracy. In this category, Fully Convolutional Network (FCN) [13] and U-Net [14] are two leading methods. Zhang *et al.* [15] use FCN-8S for segmentation of polyp candidates and then classify candidate regions with "*Texton*" features and random forest classifier. "*Texton*" features are produced by using K-means clustering algorithm on the convolution of input patch and bank of Gabor filters for different orientations.

In some applications, polyp segmentation method is a combination of more than one CNN, called "ensemble of CNNs", to overcome the diversity of shapes in polyps and their intensities. The method proposed in [16] uses three CNNs to classify input patches. It uses the method of [7] for extracting candidate regions. After that it extracts three sets of patches around each candidate region and feed them to the corresponding CNN network. These three sets of patches are partitioned based on color, texture, temporal features and shape clues. It also calculates maximum scores of all three CNNs and fully-connected layer to classify the patches. Zhang *et al.* [17] use a trained network on natural images and fine tune the weights for classification of polyp patches. This weighted network is then used for classification of each input patch into polyp and non-polyp sets. Training of CNN is a challenging problem in medical applications because of limitations in database samples. Our smart patch selection method overcomes this difficulty in the training phase of the CNNs.

### III. PROPOSED METHOD

Our proposed polyp segmentation method contains two main stages. In the first stage, we propose candidate regions of probable polyp with FCN-8S network. Then in the second stage, we use Otsu thresholding and select the largest connected component to segment polyp regions among all candidate regions. Fig 1 Shows our proposed scheme for segmentation of colonoscopy polyps.

#### A. FCN-8S Network

FCN was first proposed in [13] for semantic segmentation. This network uses stages of convolution and pooling for creating dense feature map for input image. It creates 4096 features for input image and enlarges dense feature map by using deconvolution layer and upsampling. FCN has three versions, FCN-32S, FCN-16S and FCN-8S. FCN-32S is the simplest one that enlarges just the dense feature map by upsampling with scale of 32 to generate prediction map with the size of input images. FCN-16S uses both results of pool4 and conv7 in feature extraction phase and FCN-8S uses the results of pool3, pool4 and conv7 to generate prediction map with the size of input image. Hence, FCN-16S needs up sampling with scale of 16 and FCN-8S needs up sampling with scale of 8. Fig.2 shows all these three versions of FCN networks.

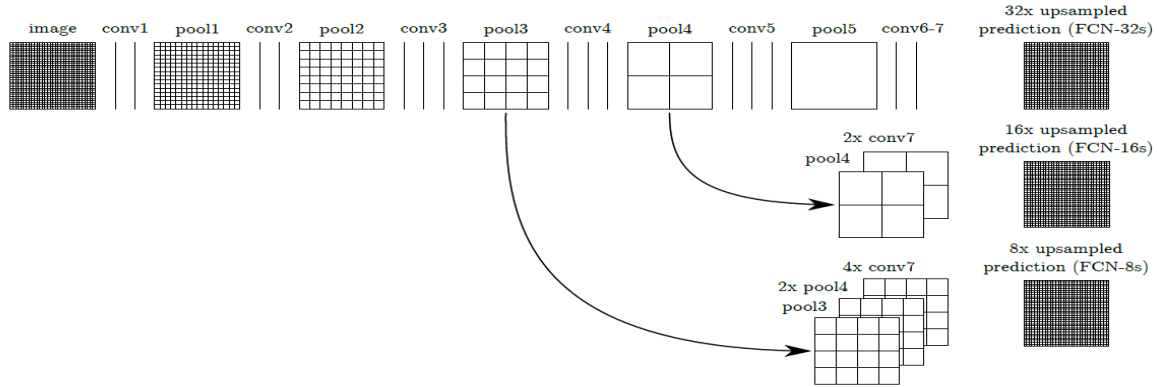

Figure 2. Different versions of an FCN [13]. FCN-8S is used in the proposed method.

Table I. Evaluation of FCN-8S for different patch selection strategies

| Training Set | Accuracy | Precision | Sensitivity | Specificity | Dice Score | FPPF |
|---|---|---|---|---|---|---|
| Image + Patch Selection Method | 0.967 | 0.836 | 0.632 | 0.991 | 0.720 | 0.10 |
| Image + Rotation + Patches of Polyps | 0.954 | 0.624 | **0.757** | 0.968 | 0.684 | 0.20 |
| Image + Rotation + Patch Selection Method | **0.977** | **0.883** | 0.748 | **0.993** | **0.810** | **0.08** |

In this paper we use FCN-8S for segmentation of polyp regions in colonoscopy images. The FCN segments candidate regions based on extracted features. Then, we use post processes to decrease false positive rate in results of FCN-8S. In the first stage of post-processing we use Otsu thresholding method to change probability map resulted from FCN-8S into a binary image and then find the largest connected component and consider it as the most probable location of polyp in the colonoscopy image.

*B. Patch Selection and Data Augmentation*

Training of FCN-8S is an important problem especially in case of medical images because of limitations in available data and ground truth set. However, one of the challenges is the insufficient number of data for training. In this paper, different tricks of augmentation will be proposed later for increasing training data for better generalization in training phase. Augmentation is very important for polyp segmentation because of the variations in polyp shapes and intensities in different images.

We employ data augmentation methods that contain image rotation or patch selection in the image, similar to the method we used in [19]. Rotation helps FCN learn different structures of polyps in different images. Our proposed patch selection method also intelligently selects center of patches from all image regions containing inside of polyps, regions of polyp borders and background regions.

IV. EXPERIMENTAL RESULTS

We use CVC-ColonDB database [18] to evaluate our proposed method. CVC-ColonDB database contains 300 images all with polyps of different shapes. These images are annotated by physicians which we use as ground truth set for evaluation of our proposed polyp segmentation method. It contains 15 sequences where each of them is from a distinct study and has the resolution of 500×574 in RGB color space.

We trained the FCN-8S on Caffe [20] using data augmentation to enlarge dataset for better training of FCN-8S. We selected patches of size 100×100 from inside, background, and boundary of polyp regions. We trained our classifier with 10 degrees of rotation between 0 and 290 and extracted 15 patches from each rotated image. Using more augmented data with more rotation steps would not increase the training accuracy and would only increase training complexities. Six patches of 15 are selected with centers inside the polyp, four with centers in the background and the other five patches with centers in the boundary of polyps. We also randomly flipped some selected patches for more generalization of learning process. Our training database contains 200 images randomly selected from original database and we left 100 remaining images for the test phase. The number of train and test images is similar to other works in the literature.

Fig 3 shows our proposed method for segmentation of polyps in colonoscopy images considering their corresponding ground truth set. We also reported dice score of segmentation results for each image after post-processing and original output data of FCN-8S. We reported quality assessment results and False Positive rate Per Frame (FPPF) for FCN-8S in different conditions of training data and selecting the largest connected component in Table I which proves that wise selection of patches will increase performance of the proposed method. We performed random patch selection and results are presented in Table I.

The first row of Table I shows the results of using just original images and the proposed patch selection method.

| Input Image | FCN-8S Result | Otsu Result | Largest Connected Component | Ground Truth | FCN Dice | Post process Dice |
|---|---|---|---|---|---|---|
| 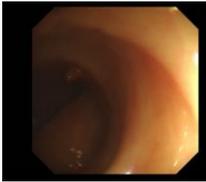 | 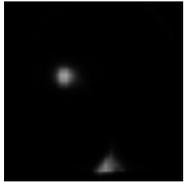 | 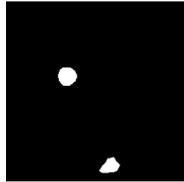 | 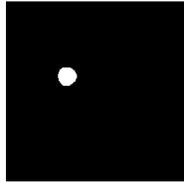 | 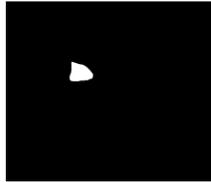 | 0.455 | 0.616 |
| 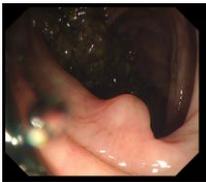 | 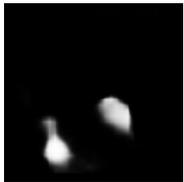 | 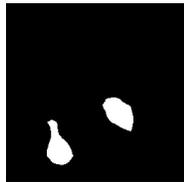 | 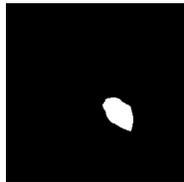 | 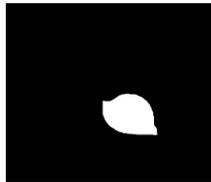 | 0.467 | 0.609 |
| 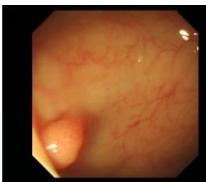 | 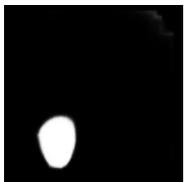 | 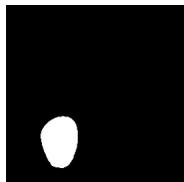 | 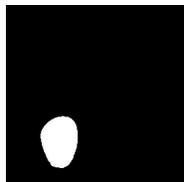 | 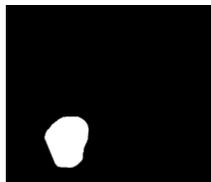 | 0.966 | 0.966 |

Figure 3. Results of the proposed segmentation method with corresponding Dice scores. Post processing has dramatically increases the Dice score.

Next row demonstrates the results of using original image and rotation with the patches of polyps that train the FCN with more sensitivity on polyp. Last row proves that wise selection of patches increases the performance of the segmentation in company with rotation. We also compared our proposed method with the method of [15] in Table II and with the methods of [3] and [8] in Table III. Our proposed method achieves higher values of sensitivity with lower FPPF in comparison with [3] and [8] and also better quality assessment results in comparison with the method proposed in [15].

Table II. Comparison of proposed method with the method of [15]

| Criterion | Accuracy | Specificity | Dice | Sensitivity |
|---|---|---|---|---|
| [15] | 0.975 | 0.988 | 0.701 | **0.757** |
| **Proposed** | **0.977** | **0.993** | **0.810** | 0.748 |

Table III. Comparison of the proposed method with [3] and [8]

| Criterion | Sensitivity | FPPF |
|---|---|---|
| Bernal et al. [3] | 67.60% | 0.300 |
| Tajbakhsh et al. [8] | 73.30% | 0.100 |
| **Proposed** | **74.80%** | **0.080** |

## V. CONCLUSION

In this paper we proposed a novel polyp segmentation method based on convolutional neural network and Otsu thresholding. We also used a wise method of patch selection for improving training phase of convolutional neural network. FCN motivated us because of its powerful ability in semantic segmentation and we used it with the Caffe framework in implementing of FCN. We evaluated our proposed method with different training sets of CVC-ColonDB database and also evaluated it for polyp segmentation. Our proposed method achieves 81% of dice score in this database which outperforms previous methods in segmentation of colorectal polyps.